\documentclass[universe,review,submit,pdftex,moreauthors]{Definitions/mdpi}
\preto{\abstractkeywords}{\nolinenumbers}
\usepackage{color} 
\usepackage[figuresright]{rotating}
\usepackage{changes}
\definechangesauthor[name={linh}, color=red]{r3}
\definechangesauthor[name={lh}, color=green]{r1}
\firstpage{1} 
\pubvolume{1}
\issuenum{1}
\articlenumber{0}
\pubyear{2022}
\copyrightyear{2022}
\datereceived{} 
\dateaccepted{} 
\datepublished{} 
\hreflink{https://doi.org/} 

\Title{Probing diversity of type II supernovae with the Chinese Space Station Telescope}

\TitleCitation{Diversity of SNe II}

\Author{Han Lin $^{1,2}$, Jujia Zhang $^{1,2}$* and Xinghan Zhang $^{3}$}

\AuthorNames{Han Lin, Jujia Zhang and Xinghan Zhang}

\AuthorCitation{Lin, H.; Zhang, J.; Zhang, X.}

\address{%
$^{1}$ \quad Yunnan Observatories, Chinese Academy of Science, Kunming 650216, PR China \\
$^{2}$ \quad Key Laboratory for the Structure and Evolution of Celestial Objects, Chinese Academy of Science, Kunming 650216, PR China \\
$^{3}$ \quad School of Physics and Information Engineering, Jiangsu Second Normal University, Nanjing, Jiangsu 211200, PR China
}

\corres{Correspondence: jujia@ynao.ac.cn}

\abstract{Type II supernovae (SNe II), which show abundant hydrogen in their spectra, belong to a class of SNe with diverse observed properties. It is commonly accepted that SNe II are produced by core collapse and explosion of massive stars. 
However, the large photometric and spectroscopic diversity of SNe II, and the mechanisms responsible for these diversities, have not been thoroughly understood.
In this review, we first briefly introduce the optical characteristics and possible progenitors of each subtype of SNe II. We then highlight the role of the Chinese Space Station Telescope in future SN studies.
With a deep limiting magnitude, the main survey project could detect SN IIP-like objects as distant as $z\sim 1.2$, and obtain UV-optical follow-up for peculiar transients, especially those long-lived events.
With a high resolution and a large field of view, the main survey camera is powerful in linking a nearby SN with its progenitor, while the integral field spectrograph is powerful in revealing the SN environment.
All this information has the potential to help enrich our understanding of supernova physics.
}


\begin{document}


\section{Introduction}
Supernovae are huge explosions that take place at the end of the evolution of stars. 
They show large diversities and can be divided into different subclasses based on their observational characteristics. 
The first-order classification of supernovae is based on whether hydrogen lines are shown in their spectra. Those that do not contain hydrogen in their spectra are classified as type I supernovae (SNe I), while those that show clear hydrogen are divided into a class of Type II supernovae (SNe II)\cite{1941PASP...53..224M}. 
With more and more supernovae have been detected by modern techniques, the classification scheme was extended according to additional spectral or photometrical properties. SNe I were divided into SNe Ia, SNe Ib, and SNe Ic, while SNe II were divided into more subcategories, i.e., SNe IIP, SNe IIL, SNe IIb, SNe IIn, and SN 1987A-like objects\cite{1997ARA&A..35..309F,2003LNP...598...21T}.
The difference in observational properties reflects the difference in stellar state (e.g., type, mass, metallicity, rotation, whether in binary system, etc.) and circumstellar environment of the progenitor star\cite{1992ApJ...391..246P,2000ApJ...544.1016H,2001ASSL..264..199C,2004MNRAS.353...87E,2004A&A...425..649H}.

It is commonly accepted that SNe Ia come from the thermonuclear explosions of white dwarfs\cite{1973ApJ...186.1007W} while SNe Ib, SNe Ic and SNe II are produced by core collapse and explosion of massive stars ($\geq$ 8-10M$_{\odot}$)\cite{1986ARA&A..24..205W,2003ApJ...591..288H,2009ApJ...691.1348A}. 
However, the exact connection between a supernova and its progenitor has not been understood thoroughly. Therefore, an important topic in the study of supernovae is to map the observed SN type to the progenitor type and to test the theory of star evolution meanwhile. 
Several indirect ways based on the analysis of the resultant supernovae are used to derive information on progenitor stars. Modeling of the nebular-phase spectra can be used to constrain the progenitor masses since spectra at this phase permit a direct view into the inner layers and contain information of nucleosynthesis yields which differ in different progenitors\cite{2012MNRAS.420.3451M,2012A&A...546A..28J,2018MNRAS.475..277J}.
Besides, Many efforts have been made for modeling the light curve, which can be used to estimate the explosion energy, ejecta mass, progenitor radius, and the mass of nickel produced by the explosion\cite{1971Ap&SS..10...28G,1977ApJS...33..515F,1980ApJ...237..541A,1985SvAL...11..145L,1991SvAL...17..210C,1993A&A...274..775B,1993ApJ...414..712P,2009ApJ...703.2205K,2011ApJ...741...41P,2012MNRAS.424.2139D,2015ApJ...814...63M}.
However, the most reliable way is to observe the progenitor itself directly. This can be done by searching the SN progenitor at pre-SN images in archives of imaging surveys.
Such a method is successful for a few of SNe, e.g., 1987A\cite{1987Natur.327...36W,1987Natur.328..318G,1989A&A...219..229W}, SN IIb 1993J\cite{1994AJ....107..662A,2004Natur.427..129M}, and a dozen of SNe IIP\cite{2009ARA&A..47...63S} and so on.
However, SNe with conclusive progenitors detected are still rare, and sometimes an SN site may not be covered due to the small Field of View (FoV) of recent space telescope. 
The wield-field space telescope, e.g., the 2-m Chinese Space Station Telescope (CSST), which has a resolution of 0.15$^{\prime \prime}$ and a large FoV of 1.1 deg${^2}$, is therefore expected to be powerful in identifying nearby progenitors.

In sections 2 and 3, we will briefly introduce the observational properties and possible progenitors of each subtype of SNe II, separately. 
In section 4, we summarize the roles of CSST in revealing the diversity of SNe II. We conclude in section 5.

\section{Observational Diversity of SNe II}
\subsection{SN IIP}
\begin{figure}[H]
	\centering
	\includegraphics[width=12 cm]{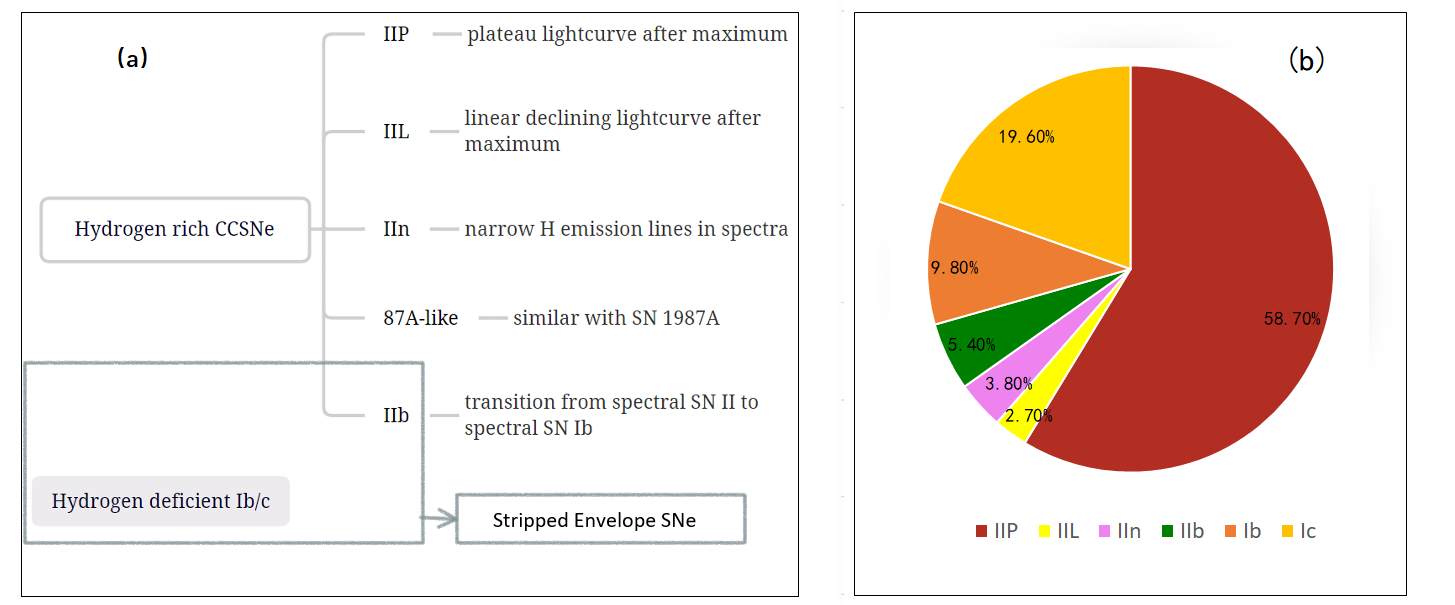}
	\caption{Panel $a$: Subtypes of SNe II and defining properties of each subtype. Panel $b$: Relative frequency of SNe II subtypes\cite{2009ARA&A..47...63S}
		\label{IItype}}
\end{figure}

The most common type of SNe II is Type IIP supernovae, which could account for nearly 60\% of the total SNe II\cite{2009MNRAS.395.1409S}. The relative ratios of each subtype are shown in Figure \ref{IItype}. SNe IIP are characterized by a plateau (the luminosity declines very slowly) of $\sim$100 days after the maximum in their light curve\cite{2012ApJ...756L..30A}. 
As shown in Figure \ref{lightcurve}, the V-band light curve of the prototypical SN IIP, 1999em, remains almost constant for nearly 80 days, after which the light curve displays a much steeper declination of $\sim$ 2 mag within the following 40 days\cite{2003MNRAS.338..939E}. Such a plateau stage is the consequence of the hydrogen recombination wave receding inward into the expanding SN ejecta\cite{1993ApJ...414..712P,2009ApJ...703.2205K}. Once the recombination front reaches the base of the massive hydrogen envelope, the plateau ends and the light curve shows a significant flux drop entering the radioactive decay tail. At this phase, the light curve is powered by radioactive decay of $^{56}$Ni and its decay product $^{56}$Co\cite{1996snih.book.....A}.

The spectra of SN 1999em at different phases are shown in Figure \ref{spectra}. At early phases, the spectra of SNe IIP are characterized by a blackbody continuum, with P-Cgyni profile of Balmer series and He~{\sc i} $\lambda$5876 supposed on it. With rapid-response transient surveys, spectroscopy can be obtained in just a few days or even hours after the explosion. Narrow emission lines of ionized species, which disappear quickly and are called as Flash-Ionized (FI) features\cite{2016ApJ...818....3K}, can be caught\cite{2014Natur.509..471G,2015MNRAS.449.1876S,2015ApJ...806..213S,2017NatPh..13..510Y,2018ApJ...859...78N,2018ApJ...861...63H,2020MNRAS.498...84Z,2022ApJ...926...20T} (see the spectra of 2013fs in Figure \ref{spectra}). 
As the ejecta expands and the photosphere recedes (in mass), material in deeper layers is exposed. At photospheric phases, spectra are formed by a series of radiative-transfer processes, i.e., photon emissions, absorptions, and scatterings\cite{1990sjws.conf..149J,2001astro.ph.11573B}. The spectra show the P-Cgyni profile of different species, such as Fe~{\sc ii} and Na I D. As the ejecta becomes totally transparent to optical photons, the spectra evolve to the nebular phase. Optical lines are mainly formed by recombination, collisional excitation, and fluorescence\cite{1992ApJ...390..602K}. The characteristic line profile is emission lines that peak near the rest wavelength\cite{2001astro.ph.11573B}.

\subsection{SN IIL}
\begin{figure}[H]
	\centering
	\includegraphics[width=10.5 cm]{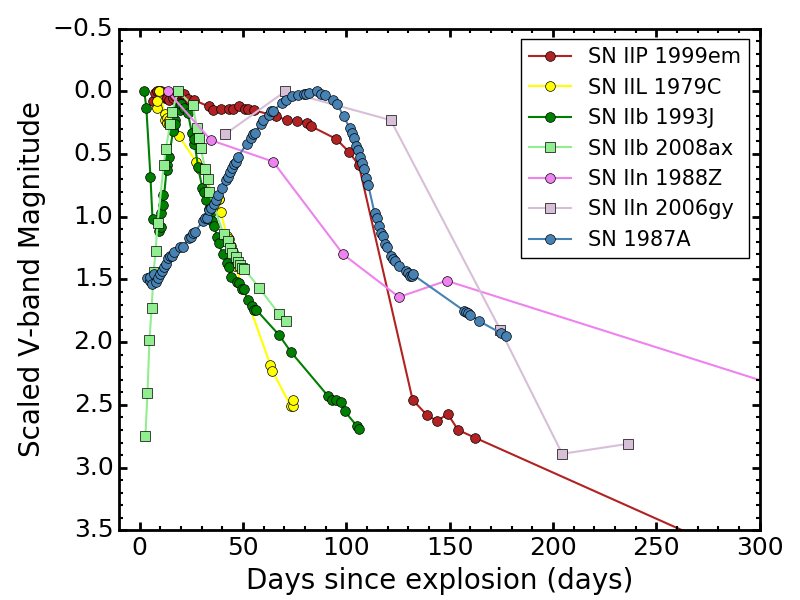}
	\caption{V-band light curves of typical individuals of each SN II subtype, including SN IIP 1999em\cite{2002PASP..114...35L,2003MNRAS.338..939E}, SN IIL 1979C\cite{1981PASP...93...36D}, SN IIb 1993J\cite{1994AJ....107.1022R}, SN IIb 2008ax\cite{2008MNRAS.389..955P}, SN IIn 1988Z\cite{1991MNRAS.250..786S,1993MNRAS.262..128T}, SN IIn2006gy\cite{2009ApJ...691.1348A} and SN 1987A\cite{1988AJ.....95...63H}. Magnitudes are normalized to peak SN magnitudes for comparisons. 
	\label{lightcurve}}
\end{figure}

SNe IIL is a rare subclass and in general, they are brighter than SNe IIP. Different from SN IIP, the light curve of SN IIL declines linearly after maximum\cite{1979A&A....72..287B}. The absorptions of H$\alpha$ in the spectra of SNe IIL are shallower\cite{1994A&A...282..731P,1996AJ....111.1660S,2014ApJ...786L..15G} (see the light curve and spectra of SN 1979C in Figure \ref{lightcurve} and Figure \ref{spectra}). Both theoretical and observational evidence suggests that the hydrogen envelope is a main factor to affect the observed properties of SNe IIP/L. The progenitor of an SN IIL is supposed to retain a lower mass of hydrogen envelope at the time of the explosion so that they cannot sustain a plateau in its light curve\cite{1983Ap&SS..89...89L,1993ApJ...414..712P,2014ApJ...786...67A,2017ApJ...850...90G}. 
Besides, the interaction between SN ejecta and circumstellar material (CSM) could produce SNe IIL-like properties as well, e.g., a fast-declining light curve, and a weak or absent H$\alpha$ absorption during the recombination phase\cite{2019A&A...631A...8H}.
However, there are no exact classification criteria for SNe IIL and SNe IIP, and whether they form two distinct populations is still debated\cite{2012ApJ...756L..30A,2014MNRAS.442..844F,2014MNRAS.445..554F,2014ApJ...786...67A,2015ApJ...799..208S,2016MNRAS.459.3939V,2017ApJ...850...90G,2018MNRAS.476.4592D}. 

\subsection{SN IIb}
\begin{figure}[H]
	\centering
	\includegraphics[width=14 cm]{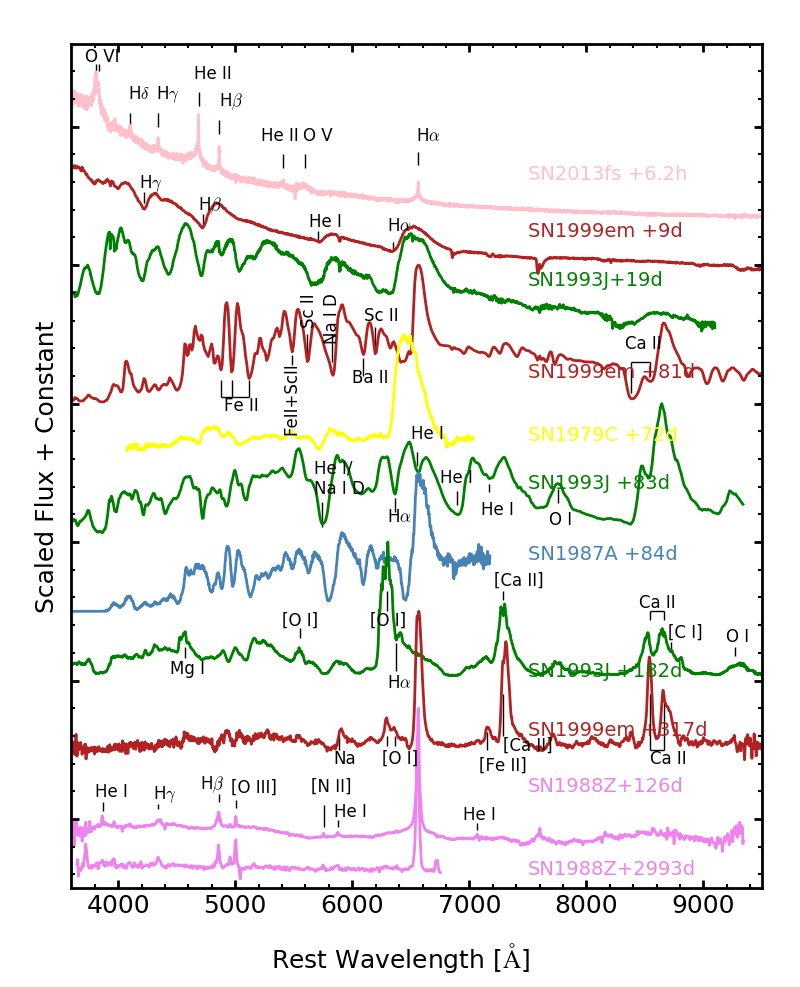}
	\caption{Optical spectra of SN IIP 2013fs\cite{2017NatPh..13..510Y}, SN IIP 1999em\cite{2001ApJ...558..615H,2002PASP..114...35L}, SN IIL 1979C\cite{1981ApJ...244..780B}, SN IIb 1993J\cite{1994AJ....108.2220F,2000AJ....120.1487M}, SN IIn 1988Z\cite{1993MNRAS.262..128T,1999MNRAS.309..343A} and SN 1987A\cite{1995ApJS...99..223P}, shifted vertically for clarity. The numbers on the right side mark the epochs since SNe explosion.
	\label{spectra}}
\end{figure}

The spectra of SN IIb show an evolution from that of SN II to that of SN Ib\cite{1988AJ.....96.1941F,1993ApJ...415L.103F}. At first, the spectra of SNe IIb show clear hydrogen lines. However, Hydrogen lines weaken at later phases and Helium lines become dominant (See the spectra evolution of SN 1993J in Figure \ref{spectra}), resembling the spectra of SN Ib. 
Such spectral transition is associated with the progenitors which are partially stripped of their hydrogen enveloped.
SN IIb, together with hydrogen deficient SN Ib\cite{1985ApJ...294L..17W} (the progenitor is supposed to lose all its hydrogen envelope) and SN Ic\cite{1995ApJ...450L..11F} (hydrogen deficient and helium deficient, the progenitor loses both H and He envelope), are called stripped envelope supernovae (SESNe)\cite{1996ApJ...459..547C}.

The light curves of part of SNe IIb, e.g., SN 1993J, show two peaks\cite{1994AJ....107.1022R}. The first peak is supposed to be in the shock cooling phase while the second peak is powered by $^{56}$Ni decay chain\cite{1993ApJ...417L..71W}. A low mass but extended envelope is required for the initial peak. Such a density structure is suggested to be produced in a binary system with the progenitor losing most of its hydrogen envelope through mass transfer to a companion star\cite{1993Natur.364..507N,1993Natur.364..509P,1994ApJ...429..300W}. 
Another part of SNe IIb, e.g., SN 2008ax, shows only one peak in their optical light curves\cite{2008MNRAS.389..955P,2011MNRAS.413.2140T}. 
The lack of shock cooling phase points to a compact progenitor. 
In conclusion, as proposed by Chevalier and Soderberg, there may exist two subtypes of SNe IIb, i.e., cIIb and eIIb\cite{2010ApJ...711L..40C}. 

\subsection{SN IIn}
Massive progenitors of SNe II lose mass and then produce CSM by means of stellar winds, impulsive eruptions, or mass transfer between interacting binaries\cite{2014ARA&A..52..487S}. 
As SN explodes in CSM, hydrodynamic and radiative interactions between SN ejecta and CSM will lead to distinct observable properties, e.g., hydrogen emission lines\cite{1982ApJ...259..302C,2003LNP...598..171C}. 
Therefore, SNe II with narrow (or intermediate-width) emission lines of hydrogen in spectra (see the spectra of SN 1988Z) are classified as SNe IIn\cite{1990MNRAS.244..269S}. 
In fact, SN IIn as a group show large diversities, since parameters associated with CSM, e.g., density, geometric distribution, and CSM composition, are needed to be considered besides the explosion parameters of their progenitors\cite{2017hsn..book..403S}.

\subsection{SN 1987A-like}
SN 1987A, with many peculiar properties, was one of the most well-studied SNe objects (see the review of \cite{1989ARA&A..27..629A} and reference therein). 
Its light curve shows a long rise ($\sim$ 90 days) before coming to the primary maximum\cite{1988AJ.....95...63H}.
The rapid initial decline of the bolometric light curve indicated a relatively compact progenitor, while the long rise to the second peak is mainly powered by slower diffusion of radioactive decay energy\cite{1988ApJ...330..218W}.
Several objects with similar long-rising light curves have since been identified and classified as 87A-like objects\cite{2005MNRAS.360..950P,2012A&A...537A.141P,2012A&A...537A.140T,2016A&A...588A...5T,2019ApJ...882L..15S,2023MNRAS.tmp..342X}. Most of the 87A-like events appear in slightly low-metallicity environments\cite{2013A&A...558A.143T}.
This type is intrinsically rare and consists of less than 3\% of core collapse Supernovae (CCSNe)\cite{2009MNRAS.395.1409S}.

\section{Progenitors of core collapse supernovae}
\begin{figure}[H]
	\centering
	\includegraphics[width=10.5 cm]{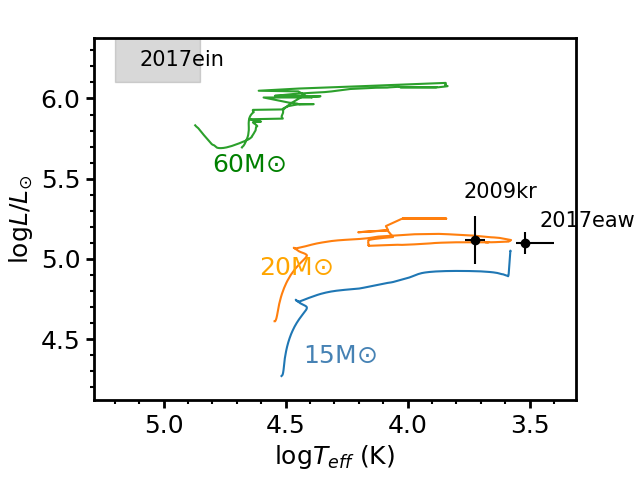}
	\caption{Hertzsprung–Russell diagram showing the locus of the progenitor of SN IIP 2017eaw\cite{2019ApJ...875..136V}, SN IIL 2009kr\cite{2010ApJ...714L.254E} and SN Ic 2017ein (gray shadow)\cite{2018ApJ...860...90V,2019ApJ...871..176X}. Blue, oragne, and green solid lines show the single-star evolutionary tracks for 15M$_{\odot}$ (with rotation at $\Omega / \Omega _{\rm crit}=0.3$), 20M$_{\odot}$ ($\Omega / \Omega _{\rm crit}=0.4$) and 60M$_{\odot}$ ($\Omega / \Omega _{\rm crit}=0.4$) at solar metallicity from Geneva model, respectively\cite{2013A&A...553A..24G,2022MNRAS.511.2814Y}.\label{hrdiagram}
    }
\end{figure}

The identifications of the progenitors of SNe provide direct information on their explosion mechanisms, which is a key point in SN studies. 
Detections of progenitors at pre-SN images suggested that SNe IIP come from red supergiant (RSG) with initial mass of 8-$\sim$17${\rm M}_{\odot}$ \cite{2009ARA&A..47...63S,2003PASP..115.1289V,2005MNRAS.359..906H,2004ApJ...615L.113M,2005PASP..117..121L,2005MNRAS.364L..33M,2006ApJ...641.1060L,2021A&A...645L...7O,2012ApJ...756..131V,2013MNRAS.431L.102M,2018MNRAS.481.2536K,2019ApJ...875..136V}. 
For instance, the effective temperature and luminosity of the progenitor of SN IIP 2017eaw are most consistent with the endpoint of 15M$_{\odot}$ track (see Figure \ref{hrdiagram}) \cite{2019ApJ...875..136V}.
For SNe IIL, observations of progenitors are rare and analysis results have significant uncertainties. A luminous yellow supergiant (YSG) with initial mass $\sim$18-24 M$_{\odot}$ could be the progenitor of SN IIL 2009kr (see Figure \ref{hrdiagram}) \cite{2010ApJ...714L.254E}. However, it is difficult to determine if the object is a single star, a binary system, or a compact cluster\cite{2010ApJ...714L.280F,2015MNRAS.447.3207M}. The progenitor of SN 2009hd could be a RSG or YSG with an initial mass smaller than 20M$_{\odot}$\cite{2011ApJ...742....6E}. In conclusion, the progenitors of SNe IIL and their mass range are still in debate.

Stripped envelope SNe IIb, whose progenitor stars are supposed to be partially stripped of their hydrogen envelope, may arise from two different channels. For SN IIb 1993J, the detection of the signatures of the companion star ten years after SN 1993J exploded supported the binary scenario of a pair of K-type supergiant primary star (${\rm M}_{ZAMS} \approx 15{\rm M}_{\odot}$) plus B-type supergiant companion star (${\rm M}_{ZAMS} \approx 14 {\rm M}_{\odot}$)\cite{2004Natur.427..129M}. 
On the other hand, Crockett et al. found that both a single Wolf-Rayet (WR) star and an interacting binary are possible to be the progenitor of SN 2008ax\cite{2008MNRAS.391L...5C}. However, Folatelli et al. suggested a single WR star was not compatible with the new, revisited photometry\cite{2015ApJ...811..147F}.

SNe Ib/c, which are cousins of SNe IIb, may be produced by multiple channels as well.
Theoretically, Star above 25-30M$_{\odot}$ is supposed to evolve to a WR star, lose its hydrogen (and helium) envelope by stellar winds, and then explode as SN Ib/c\cite{1986ApJ...306L..77G}. Such a scenario is supported by the evidence that SNe Ib/c trace the star formation of their host galaxies more accurately than SNe II\cite{2008MNRAS.390.1527A}. However, smaller ejecta masses for stripped envelope SNe are inconsistent with those expected from very massive stars\cite{2016MNRAS.457..328L}. Alternatively, it has been suggested that stripped-envelope SNe may come from moderately massive progenitors (8$\sim$20M$_{\odot}$) in binaries. In such a channel, the progenitor could shed its hydrogen (and helium) envelope through interaction with a binary companion\cite{1992ApJ...391..246P,1995PhR...256..173N}. It is suggested that a large ratio of massive stars (more than 70\%) could be in binaries\cite{2012Sci...337..444S}.  
The explosion rate of SN Ib/c, which is higher than the single WR stars channel expected\cite{2011MNRAS.412.1522S}, can be explained with binary population models\cite{2013MNRAS.436..774E} as well.
Thus, binary systems could constitute a significant fraction of SNe Ib/c progenitors. Successful detections of progenitor for SN Ib/c are rare. 
For SNe Ib, the late time observations (+740 d) of iPTF13bvn favor a helium giant progenitor in binary\cite{2016MNRAS.461L.117E} rather than a WR star\cite{2013ApJ...775L...7C}. 
SN 2019yvr was the second SN Ib with progenitor detected, and its progenitor could be in a binary system as well\cite{2021MNRAS.504.2073K,2022MNRAS.510.3701S}.
For SN Ic, the progenitor candidate of SN 2017ein found in archival Hubble Space Telescope (HST) images could be the first time that a progenitor candidate of SN Ic has ever been identified. The very blue and hot candidate would have to be very massive (see Figure \ref{hrdiagram}) \cite{2018ApJ...860...90V,2019ApJ...871..176X}.

Luminous Blue Variable stars (LBV) are the possible progenitor candidates for the majority of SNe IIn since giant eruptions of LBV can produce CSM which are dense enough 
\cite{2017hsn..book..403S}. Gal-Yam et al. identified a luminous point source, which is likely a LBV, as the possible progenitor of SN 2005gl\cite{2007ApJ...656..372G}. 
A series of LBV-like giant eruptions was also observed for the progenitor of SN 2009ip\cite{2010AJ....139.1451S,2013MNRAS.430.1801M,2013ApJ...767....1P}. 
However, we must notice that SNe IIn are defined by external features associated with circumstellar interaction and thus, they can not be linked to a singular progenitor type.

The Progenitor of SN 1987A, known as Sk-69$^{\circ}$202, was detected in archival images. It was a compact blue supergiant (BSG) with a ZAMS mass of 14-20M$_{\odot}$\cite{1987Natur.327...36W,1989A&A...219..229W,1987Natur.328..318G,1992PASP..104..717P}. It is surprising that Sk-69$^{\circ}$202 was not a RSG when it exploded. Besides, Nitrogen-enhanced circumstellar material with a triple-ring structure\cite{1996ApJ...464..924L} was supposed to be ejected from the progenitor of SN 1987A about 20,000 years before the explosion\cite{2000ApJ...528..426C}. Many theoretical models, such as mass transfer in binaries, binary merger, a rapidly rotating single-star, extensive mass loss et al., are proposed to explain why the progenitor of SN 1987A ends as a compact BSG\cite{1989ARA&A..27..629A,1992PASP..104..717P,1988ApJ...331..388S,1987Natur.327..597H,1988ApJ...324..466W,2017MNRAS.469.4649M,2021ApJ...914....4U}. 
However, progenitors of 87A-like SNe are not settled yet and this type has much to be learned about.

Linking the types and mass ranges of progenitor stars with the characteristic of SN is one of the major goals in studies of SNe explosions. Since late 1990s, efforts have been made in systematic search for progenitor stars of all nearby SNe.
However, we still do not have a complete picture about the progenitor channels of core collapse supernovae. RSG with an initial mass between 8-17M$_{\odot}$ was confirmed to be the progenitor of SNe IIP.
However, the reason for the "red supergiant problem", that lacking detected RSG progenitors with initial masses between 17–25M$_{\odot}$ is still unclear.
High mass RSG ($M_{ZAMS} \geq 17{\rm M}_{\odot}$) with sufficiently low metallicity could explod as SN IIP, however such progenitors in low luminosity, low metallicity galaxies are not easy to be detected\cite{2018NatAs...2..574A}.
Inadequate bolometric correction\cite{2018MNRAS.474.2116D}, additional extinction resulting from the dust\cite{2012MNRAS.419.2054W}, limited wavelength coverage\cite{2019ApJ...875..136V}, and so on., all of these would underestimate the luminosity, and thus the initial mass of progenitor stars.
On the other hand, Smartt proposed that stars in the mass range between 8-17M$_{\odot}$ might collapse to form black holes with failed SNe as theoretically expected\cite{2016ApJ...821...38S}. Such a scenario was supported by Adams et al. since they identified a failed supernovae candidate, N6946-BH1, a $\sim$25M$_{\odot}$ red supergiant experienced a weak optical outburst\cite{2017MNRAS.468.4968A} and has since disappeared in the optical\cite{2021MNRAS.508.1156B}. Its bolometric luminosity is fading as $\sim t^{-4/3}$, consistent with the models of fallback accretion onto a black hole as well\cite{2017MNRAS.468.4968A}.

For other core collapse supernovae, e.g., SNe IIL, IIb and IIn, there are only a few of progenitor detections, which are insufficient to draw conclusions for the whole class. 
The progenitors of CCSNe are suggested to experience larger amount of mass-loss with the sequence of SNe IIP–IIL–IIb–Ib/c, since hydrogen envelope is decreasing with this sequence. Mass loss, rotation, metallicity and binary interaction, all of these factors will influence the evolution of massive stars and thus, the observed properties of SNe\cite{1995PhR...256..173N,2003ApJ...591..288H}. 
In conclusion, the mapping of massive stars (especially those above 17${\rm M}_{\odot}$) with their death product, as well as the role of rotation, metallicity, mass loss and binary interaction are still needed to be determined, while only a very limited number of clear progenitor detection.

\section{Roles of CSST in SNe II studies}

\subsection{Chinese Space Station Telescope}
The 2-m Chinese Space Station Telescope (CSST), which will be launched at the end of 2023, is expected to be a powerful tool in revealing the diversity of SNe II. 
The main survey project, which is capable of performing both photometric imaging and spectroscopic slitless
surveys, will cover a survey area of 17500 deg$^2$ in ten years and obtain high-quality photometric and spectral data for hundreds of millions of targets. Its photometric survey, with an average limiting magnitude of 25.5 mag, is designed to cover a wavelength range of 255-1000nm in NUV, u, g, r, i, z, and y bands, while the spectroscopic survey will cover GU, GV, and GI bands down to a limiting magnitude of about 21 mag\cite{2011SSPMA..41.1441Z,2019ApJ...883..203G}.
Besides the multi-color imaging and slitless spectroscopy survey camera, a multichannel imager (MCI), an integral field spectrograph (IFS), a cool planet imaging coronagraph, and a terahertz receiver are equipped in CSST as well.
With FoV of $7.68^{\prime} \times 7.68^{\prime}$ and limiting magnitude of $29 \sim 30$ mag, MCI is designed to provide flux calibration for the main survey, as well as UV-optical ultradeep field observations (Observations in three bands, 250-400nm, 400-700nm, and 700-1100nm could be performed simultaneously). IFS is designed to obtain two-dimensional spectra that contain both spatial and spectral information. 
Designed parameters of the main survey camera, MCI, and IFS are listed in Table~\ref{csst_para}.

\subsection{CSST for mapping SNe II with progenitors}
With an average limit magnitude of 25.5mag, the photometric survey could probe the brightest RSG ($L \sim 5.4L_{\odot}$, $M_V \sim -7$ mag \cite{2022arXiv221114147M}) within $\sim$30Mpc. With a resolution of 0.15$^{\prime \prime}$, extra information such as color and spectral energy distribution (SED), is needed to distinguish progenitor stars and clusters in some cases. Such information could be obtained by CSST as well.
If a SN is located with a coeval cluster, SED and multiband photometry could be used for studying the stellar population, thus the turn-off mass, which provides clues of the initial mass of the progenitor, could be estimated\cite{2004ApJ...615L.113M}.
Besides, deep and high-resolution images taken after SNe fading away are also critical to distinguish between different progenitor scenarios. 

Consider the local rate of CCSNe\cite{1999A&A...351..459C} and the average luminosity density of galaxies\cite{2001MNRAS.324..825C}, $6.7^{+5.7}_{-4.1} {year}^{-1}$ CCSNe will appear within 30Mpc. Such a rate is consistent with the volume rate calculated from Lick Observatory Supernova Search (LOSS)\cite{2011MNRAS.412.1473L} and the number of CCSNe (i.e., 92) found within 28 Mpc in a 10.5-year period. 
In these 92 CCSNe, only 26\% of them have a pre-SN image in the HST archive, partly because the SN position is missed due to the small FoV of the HST camera\cite{2009MNRAS.395.1409S}. 
With a large FoV of 1.1 deg${^2}$, which is $\sim$ 300 times of Advanced Camera for Surveys (ACS)/Wide Field Channel (WFC) of HST, 
the coverage of host galaxies, and then, the detection rate of SN progenitors, will be improved.

\subsection{CSST for revealing the environment of SNe II}
Due to the rarity of sufficient deep and high-resolution pre-images of nearby SN explosions, environment properties in the vicinity of the SN site, e.g., metallicity, star formation, and radial position, could provide additional statistical constraints for progenitor systems of different SN types. Both photometric, e.g., H$\alpha$+[N II]+R-band images, and spectroscopic observations have been used for environment analysis\cite{2006A&A...453...57J,2008MNRAS.390.1527A,2010MNRAS.407.2660A}. 
Since the ambiguities of SN classification, Anderson believes that using specific SN features to investigate environment trends in future works will significantly increase insight on SN progenitors\cite{2015PASA...32...19A}. They also strengthen the importance of wide-field integral field units (IFU) observations, which enables simultaneous spatial and spectral information for SN site. The IFS equipped in CSST, with a spatial resolution of 0.2$^{\prime \prime}$ and FoV of $6^{\prime \prime} \times 6^{\prime \prime}$, could provide a detailed environment structure within a radius of $\sim$2kpc for supernovae within 40Mpc. One of the main scientific goals of the CCST is to understand galaxy formation and evolution by scrutinizing the star formation and stellar population of nearby galaxies. Once this information is provided, it can be used to estimate SN environment properties.

\subsection{CSST for observing peculiar transients}
\begin{figure}[H]
	\centering
	\includegraphics[width=12 cm]{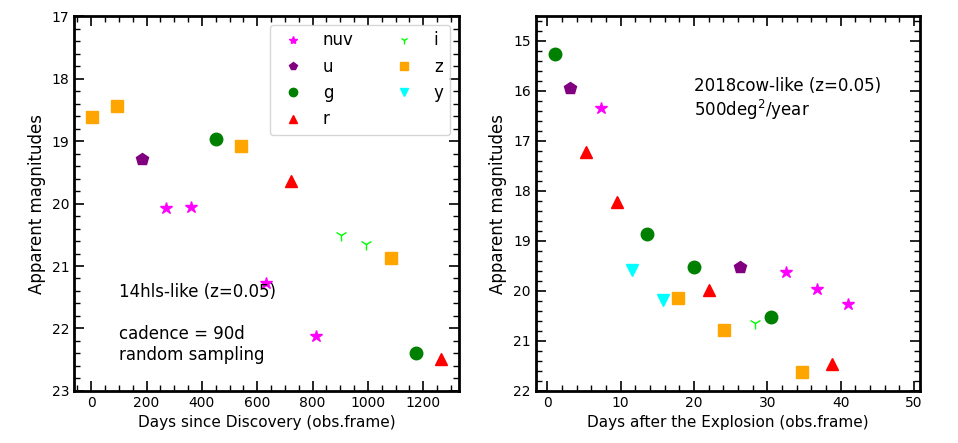}
	\caption{Simulated light curves for iPTF14hls and 2018cow with a cadence of 90 days and 3 days, separately. Different colors and shapes represent different bands. In our simulation, the target randomly falls on a certain filter in each observation.
	\label{simu_lc}}
\end{figure}

Recent surveys have revealed new classes of peculiar transients, e.g., the bright, long-lived multi-peak iPTF14hls and Fast Blue Optical Transients (FBOT). 
A series of models have been proposed for iPTF14hls\cite{2018MNRAS.477...74A,2018MNRAS.475.1198S,2018A&A...610L..10D,2020ApJ...897..156U,2020MNRAS.491.1384M} and FBOT 2018cow\cite{2019MNRAS.484.1031P,2019ApJ...872...18M,2019MNRAS.487.5618L,2019MNRAS.487.2505K,2020ApJ...903...66L}, however, their true nature is still in debate.
These peculiar transients, which pose significant challenges to conventional supernova explosion physics are critical for understanding the stellar evolution theory. 
Considering the FoV of all seven filters and the exposure time of 150s, the multi-band main survey camera will take about 90 days to cover the whole 17500 deg$^2$.
Such a survey strategy is powerful in observing the long-lived supernova, as well as pre-SN activities.
In Figure \ref{simu_lc}, we simulate the observation results (each mock sampling randomly selects a certain filter) taken by the multi-band camera for iPTF14hls-like events and 2018cow-like events. 
The bright iPTF14hls-like events at z=0.05 could be observed for years. However, a survey strategy with a high cadence (e.g., covering 500 square degrees in one year) should be conducted for those fast-evolving events.

The collapse of iron core may not be the only mechanism that a massive star can die and make a supernova. It is suggested that extremely massive stars could explode as pair-instability SNe (PISNe).
Oxygen cores of these extremely massive stars may reach to a high temperature but low density condition, in which electron–positron pairs are copiously produced. The radiation pressure is reduced, and the star contracts, leading to a runaway thermonuclear explosion to completely disrupted the star\cite{1967PhRvL..18..379B,1967ApJ...148..803R}.
Besides, massive stars may undergo pulsational pair instability and eject massive shells\cite{2007Natur.450..390W}. 
Most of these PISNe are supposed to occur in the early universe, thus, could be the target of the ultradeep field observations of MCI.
At the low mass end (8-10M$_{\odot}$) of SNe II progenitor, electron capture by $^{20}$Ne and $^{24}$Mg are supposed to trigger the collapse and produce electron-capture supernovae (ECSNe)\cite{1980PASJ...32..303M,1984ApJ...277..791N,1987ApJ...322..206N}. Several transients are ECSNe candidates\cite{2009MNRAS.398.1041B,2009ApJ...705.1364T,2021NatAs...5..903H,2013MNRAS.431.2599M,2014A&A...569A..57M} but no unequivocal observations have been obtained. 
With the survey project of CSST and ground-based telescopes, new peculiar transients may be discovered and monitored, and thus, progress will be made in stellar evolution theory.

A supernova first shines in X-ray and UV when shock breakout. After the breakout, the hot ejecta expands and cools, and the SN radiated mainly in UV and optical\cite{1978ApJ...225L.133F}. Radiation during the shock breakout and shock cooling phase contains information about the progenitor star, e.g., density structure.
Moreover, ultraviolet flux accounts for a large part of the total bolometric luminosity for SN II 
\cite{2018MNRAS.475.3959H}. 
All massive stars lose their mass with different rates and mechanisms, leading to different density structures of CSM. When CSI happened, both the shocked region and the unshocked region could radiate in the UV and optical\cite{2003LNP...598..171C}. Such emission could be used to derive the mass loss history of the progenitor star, which in turn, could help to determine the progenitor type. 
Thus, UV observations are paramount for both detecting objects with shock breakout or CSI, and estimating the bolometric luminosity especially at early phases. 
However instruments which could performed UV obervations, e.g., $Swift$ and HST, are scarce. CSST, equipped with a multi-band survey camera and MCI, could be one of the few instruments that are able to provide near-UV observations in the future.

\subsection{CSST for SNe II cosmology}
The intrinsic high luminosity and the relation between peak brightness and the shape of the light curve (e.g., the Phillips relation\cite{1993ApJ...413L.105P}) make SNe Ia the best “standardizable candle” for cosmological distance measurement. The magnitude/distance-redshift relation of SNe Ia can be used to constrain the cosmological parameters, e.g., the Hubble constant (H0), the mass density, the cosmological constant, and the deceleration parameter. Based on SN Ia cosmology, both the High-z Supernova Search Team and the Supernova Cosmology Project favored the universe model with a positive cosmological constant and a current accelerated expansion\cite{1998AJ....116.1009R,1998ApJ...507...46S,1999ApJ...517..565P}.
Using 42 Cepheid-calibrated SNe Ia, the uncertainty in the local determination of H0 has been decreased to 1.4\%\cite{2022ApJ...934L...7R}. 

However, To date, there is no explanation found for the “H0 tension”, which refers to the discrepancy between the value derived from cepheid calibrated SNe Ia and that inferred via the cosmic microwave background\cite{2022ApJ...934L...7R}.
While H0 can be measured locally, high-redshift objects are critical for distinguishing different universe models. However, only a small number of SN Ia are detected at z > 1.4 due to a long time delay between the formation of the progenitor star and the explosion of the supernova\cite{2008ApJ...681..462D}.
Other independent distance measurements are needed as much as possible. 
Although SN II is less luminous than SN Ia, its intrinsic explosion rate, is higher than SN Ia, especially at high redshift\cite{2012A&A...545A..96M}.  
Unlike the relatively homogeneous appearance of SNe Ia, SNe II exhibit a large diversity, making their standardization more difficult than that of the former. However,
several methods have been proposed using theoretical or empirical relation of SNe II to measure distance, mainly including the “Expanding Photosphere Method” (EPM)\cite{1974ApJ...193...27K}, the “Standard Candle Method” (SCM)\cite{2002ApJ...566L..63H}, and the “Photometric Color Method” (PCM)\cite{2015ApJ...815..121D}.
The dispersion in the Hubble diagram obtained for PCM and SCM are 0.44 and 0.29 mag, separately\cite{2015ApJ...815..121D}, still larger than the 0.15 mag precision yielded by SNe Ia\cite{1996AJ....112.2408H}.
New knowledge and insight about Type II supernovae and their progenitors from CSST observation are essential to improve their accuracy as distance indicators. 

\begin{figure}[H]
	\centering
	\includegraphics[width=12 cm]{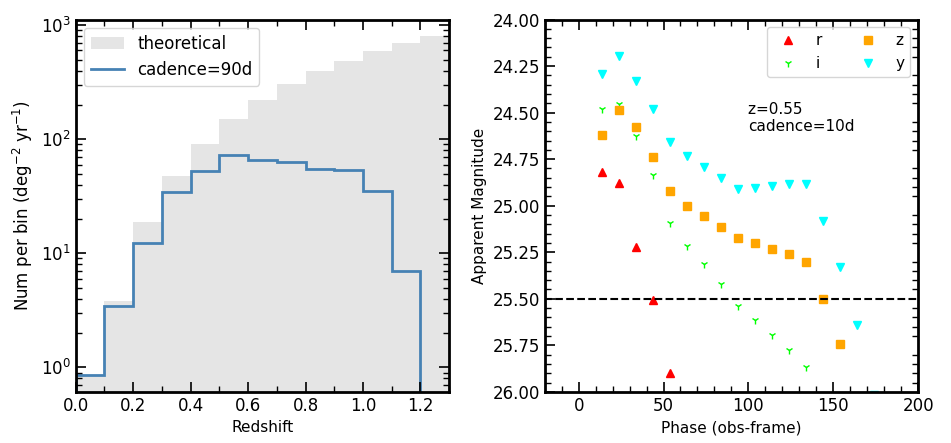}
	\caption{Panel $a$: Redshift distribution of SNe II simulated for multi-band imaging survey. Panel $b$: Simulated light curve for average SNe IIP with a regular candence of 10 day at $z=0.55$.
	\label{detect_rate}}
\end{figure}

We assess the detectability of SNe II by mock observations of the multi-band imaging survey of CSST. 
The volume rate of SNe II is calculated following the method in \cite{2012A&A...545A..96M} and reference therein. 
Since the heterogeneity of SNe II, a mean spectra energy distribution of SNe IIP\cite{1999ApJ...521...30G} ($M_{\rm B}$ scaled to -16.7 mag) is assumed for simplification. 
The observer-frame light curve is then generated using $SNcosmo$\cite{2022zndo...6363879B}. Considering the FoV and an exposure time of 150s, 
a regular cadence of 90 days is assumed for mock observations.
The result is shown in the left panel of Figure \ref{detect_rate}. In our simulations, the photometric survey could detect SN IIP at z$\sim$1.2.
However, a higher cadence (e.g., 10 days, see the right panel of Figure \ref{detect_rate}) must be conducted to obtain a relatively intact light curve that could be utilized in PCM.

\section{Conclusion}
SNe II, produced by core collapse and explosion of massive stars, are the most common type of explosion in the universe. In this review, we briefly introduce the classification scheme, the observational diversities and possible progenitor stars of SNe II, and the role that CSST can play in this field. 

Though SNe have been studied extensively for decades and remarkable achievements have been made, a host of scientific questions still exist. The basic questions are which kinds of progenitor stars explode to which kinds of SNe and the physical mechanisms that lead to the explosions. 
The main survey camera of CSST, with high resolution and large FoV, could obtain deep, high-resolution pre-SN and post-SN images, thus, is expected to be powerful in identifying progenitors and distinguishing between different progenitor scenarios. 
Moreover, spatial and spectral information of the SN environment obtained by the IFS module could provide additional constraints for progenitor systems.
Supernovae are involved in many relevant objectives of
the research
, e.g., mass loss and its influence on the evolution of massive stars, cosmology, and so on.
Well-observed individuals, especially those with peculiar properties, are critical for stellar physics, while those at a cosmological distance are possible to serve as distance indicators. 
With CSST and rapid-response surveys by ground-based telescope, a large number of transients with intense multiwavelength follow-up could be obtained and thus, help to enrich our understanding in these fields.

\vspace{6pt} 



\authorcontributions{}
H.L. is the first author who is responsible for this paper; 
J.Z. supervised/supported H.L. for this paper and simulated light curves of peculiar transients .
X.Z. gave useful comments for this paper. 
All authors have read and agreed to the published version of the manuscript.

\funding{}
This work is supported by the National Key R\&D Program of China with No. 2021YFA1600404, the National Natural Science Foundation of China (NSFC, 12173082), and science research grants from the China Manned Space Project with No. CMS- CSST-2021-A12, the Yunnan Province Foundation (202201AT070069),  the Top-notch Young Talents Program of Yunnan Province, and the Light of West China Program provided by the Chinese Academy of Sciences.

\institutionalreview{}
Not applicable.

\informedconsent{}
Not applicable.

\dataavailability{} 
Some data are obtained from the Open Supernova Catalog \\
at https://github.com/astrocatalogs, and 
the Weizmann Interactive Supernova Data Repository (WISeREP)
at https://wiserep.weizmann.ac.il/. \\
Some data used in this work are available from the published literature.

\acknowledgments{}

\conflictsofinterest{} 
The authors declare no conflict of interest. The funders had no role in the design of the study; in the collection, analyses, or interpretation of data; in the writing of the manuscript, or in the decision to publish the results.



\abbreviations{Abbreviations}{
The following abbreviations are used in this manuscript:\\
\noindent 
\begin{tabular}{@{}ll}
SN/SNe & supernova/supernovae\\
CCSNe & Core Collapse Supernovae\\
SESNe & Stripped Envelope Supernovae\\
SLSNe & Superluminous Supernovae \\
PISNe & Pair-instability Supernovae \\
ECSNe & Electron-capture Supenovae\\
CSM & Circumstellar Material \\
CSI & Circumstellar Interaction\\
RSG & Red Supergiant\\
BSG & Blue Supergiant\\
YSG & Yellow Supergiant\\
WR & Wolf-Rayet stars \\
LBV & Luminous Blue Varible \\
FBOT & Fast Blue Optical Transients     \\
FoV & Field of View	       \\
SED & spectral energy distribution      \\
FI & Flash-Ionized  \\
CSST & Chinese Space Station Telescope      \\
MCI & multichannel imager      \\
IFS & integral field spectrograph      \\
IFU & integral field units      \\
LOSS & Lick Observatory Supernova Search      \\
HST & Hubble Space Telescope     \\
ACS & Advanced Camera for Surveys      \\
WFC & Wide Field Channel \\
EPM & Expanding Photosphere Method \\
PCM & Photometric Color Method \\ 
SCM & Standard Candle Method     
     
\end{tabular}
}

\appendixtitles{no} 
\appendixstart
\appendix
\section[\appendixname~\thesection]{}

\begin{sidewaystable}[]
    \caption{Parameters of multi-color imaging and slitless spectroscopy survey, MCI and IFS. \label{csst_para}}
	\begin{tabular}{llcll}
		\cline{1-4}
		\multicolumn{4}{c}{multi-color imaging and slitless spectroscopy survey}                                                                                                                                                                                                                                                                                              &  \\ \cline{1-4}
		\multicolumn{2}{l}{FoV}                                                                                                                                                                                               & \multicolumn{2}{c}{1.1deg$^2$}                                                                                                                &  \\ \cline{1-4}
		\multicolumn{2}{l}{survey area}                                                                                                                                                                                       & \multicolumn{2}{c}{$\geq$17500${\rm deg}^2$}                                                                                                  &  \\ \cline{1-4}
		\multicolumn{2}{c|}{Photometric survey}                                                                                                                                                                               & \multicolumn{2}{c}{Spectroscopic survey}                                                                                                      &  \\ \cline{1-4}
		Bands              & \multicolumn{1}{l|}{\begin{tabular}[c]{@{}l@{}}NUV (255nm-317nm), u (322nm-396nm), \\ g (403nm-545nm), r(554nm-684nm), \\ i (695nm-833nm), z (846nm-1000nm) \\ and y(937nm-1000nm)\end{tabular}} & \multicolumn{1}{l}{Bands}               & \begin{tabular}[c]{@{}l@{}}GU (255um-420nm), \\ GV (400nm-650nm), \\ GI (620nm-1000nm)\end{tabular} &  \\ \cline{1-4}
		spatial resolution & \multicolumn{1}{l|}{$\sim$0.15$^{\prime \prime}$}                                                                                                                                                & \multicolumn{1}{l}{Spectral resolution} & R$\geq$200                                                                                          &  \\ \cline{1-4}
		limit magnitude    & \multicolumn{1}{l|}{25.5 on average (S/N$>$5, AB mag)}                                                                                                                                           & \multicolumn{1}{l}{limit magnitude}     & GU,GI 20; GV21 (S/N$>$5)                                                                            &  \\ \cline{1-4}
		\multicolumn{4}{c}{MCI}                                                                                                                                                                                                                                                                                                                                               &  \\ \cline{1-4}
		\multicolumn{2}{l}{FOV}                                                                                                                                                                                               & \multicolumn{2}{c}{$7.68^{\prime} \times 7.68^{\prime}$}                                                                                      &  \\ \cline{1-4}
		\multicolumn{2}{l}{survey area}                                                                                                                                                                                       & \multicolumn{2}{c}{$\geq$300arcmin$^2$}                                                                                                       &  \\ \cline{1-4}
		\multicolumn{2}{l}{bands}                                                                                                                                                                                             & \multicolumn{2}{c}{250nm-400nm ,400nm-700nm,700nm-1100nm}                                                                                     &  \\ \cline{1-4}
		\multicolumn{2}{l}{spatial resolution}                                                                                                                                                                                & \multicolumn{2}{c}{0.18$^{\prime \prime}$}                                                                                                    &  \\ \cline{1-4}
		\multicolumn{2}{l}{limit magnitude}                                                                                                                                                                                   & \multicolumn{2}{c}{29-30mag (SN$>$5)}                                                                                                         &  \\ \cline{1-4}
		\multicolumn{4}{c}{IFS}                                                                                                                                                                                                                                                                                                                                               &  \\ \cline{1-4}
		\multicolumn{2}{l}{FoV}                                                                                                                                                                                               & \multicolumn{2}{c}{$\geq 6^{\prime \prime} \times 6^{\prime \prime}$}                                                                         &  \\ \cline{1-4}
		\multicolumn{2}{l}{wavelength coverage}                                                                                                                                                                               & \multicolumn{2}{c}{0.35-1.0um}                                                                                                                &  \\ \cline{1-4}
		\multicolumn{2}{l}{spatial resolution}                                                                                                                                                                                & \multicolumn{2}{c}{$\leq 0.2^{\prime \prime}$}                                                                                                &  \\ \cline{1-4}
		\multicolumn{2}{l}{Spectral resolution}                                                                                                                                                                               & \multicolumn{2}{c}{R$\geq$1000}                                                                                                               &  \\ \cline{1-4}
		\multicolumn{2}{l}{limit magnitude}                                                                                                                                                                                   & \multicolumn{2}{c}{17 mag/arcsec$^2$ (B band, 200s$\times$20,SN=10)}                                                                          &  \\ \cline{1-4}
	\end{tabular}
\end{sidewaystable}

\begin{adjustwidth}{-\extralength}{0cm}

\reftitle{References}

\bibliographystyle{mdpi.bst}
\bibliography{ref.bib}

\end{adjustwidth}
\end{document}